\documentstyle[12pt,psfig]{article}
\newlength{\dinwidth}
\newlength{\dinmargin}
\begin{document}
\vspace{1 cm}
\newcommand{\Gev}       {\mbox{${\rm GeV}$}}
\newcommand{\Gevsq}     {\mbox{${\rm GeV}^2$}}
\newcommand{\qsd}       {\mbox{${Q^2}$}}
\newcommand{\x}         {\mbox{${\it x}$}}
\newcommand{\smallqsd}  {\mbox{${q^2}$}}
\newcommand{\ra}        {\mbox{$ \rightarrow $}}
\newcommand{\logxp}     {\mbox{$\ln(1/x_{p})$}}
\newcommand{\logxpmax}  {\mbox{$\ln(1/x_{p})_{max}$}}
\newcommand{\ee}  {\mbox{$e^+e^-$}}
\newcommand{\sleq} {\raisebox{-.6ex}{${\textstyle\stackrel{<}{\sim}}$}}
\newcommand{\sgeq} {\raisebox{-.6ex}{${\textstyle\stackrel{>}{\sim}}$}}
% ---- commands from paul -----
\def\ctr#1{{\it #1}\\\vspace{10pt}}
\def\si{{\rm si}}
\def\Si{{\rm Si}}
\def\Ci{{\rm Ci}}
\def\qsq{Q^{2}}
\def\yjb{y_{_{JB}}}
\def\xjb{x_{_{JB}}}
\def\qjb{\qsq_{_{JB}}}
\def\gap{\hspace{0.5cm}}
\title {
{\bf Measurement of multiplicity\\
and momentum spectra \\
in the current fragmentation region \\
of the Breit frame at HERA }}
\author{ZEUS Collaboration}
\date{ }
\maketitle
\vspace{5 cm}
\begin{abstract}
\noindent

Charged particle production has been measured in Deep Inelastic Scattering
(DIS) events using the ZEUS detector over a large range of $Q^2$ from 10
to $1280 {\rm\ GeV}^2$. The evolution with $Q$ of the charged multiplicity
and scaled momentum has been investigated in the current fragmentation
region of the Breit frame. The data are used to study QCD \linebreak
coherence effects in DIS and are compared with corresponding \ee~data in
order to test the universality of quark fragmentation.

\end{abstract}

\vspace{-20cm}
\begin{flushleft}
\tt DESY 95-007 \\
January 1995 \\
\end{flushleft}

\setcounter{page}{0}
\thispagestyle{empty}

\newpage

%   05/12/94 501170902  MEMBER NAME  AUTH019  (ZEUS)     M  TEX
%
\def\3{\ss}
\footnotesize
\renewcommand{\thepage}{\Roman{page}}
\begin{center}
\begin{large}
The ZEUS Collaboration
\end{large}
\end{center}
\noindent
M.~Derrick, D.~Krakauer, S.~Magill, D.~Mikunas, B.~Musgrave,
J.~Repond, R.~Stanek, R.L.~Talaga, H.~Zhang \\
{\it Argonne National Laboratory, Argonne, IL, USA}~$^{p}$\\[6pt]
R.~Ayad$^1$, G.~Bari, M.~Basile,
L.~Bellagamba, D.~Boscherini, A.~Bruni, G.~Bruni, P.~Bruni, G.~Cara
Romeo, G.~Castellini$^{2}$, M.~Chiarini,
L.~Cifarelli$^{3}$, F.~Cindolo, A.~Contin,
I.~Gialas, P.~Giusti, \\
G.~Iacobucci, G.~Laurenti, G.~Levi, A.~Margotti,
T.~Massam, R.~Nania, C.~Nemoz, F.~Palmonari, A.~Polini, G.~Sartorelli,
R.~Timellini, Y.~Zamora Garcia$^{1}$,
A.~Zichichi \\
{\it University and INFN Bologna, Bologna, Italy}~$^{f}$
% ...
\\[6pt]
\clearpage
\noindent
\hspace*{1mm}
$^{ 1}$ supported by Worldlab, Lausanne, Switzerland \\
\hspace*{1mm}
$^{ 2}$ also at IROE Florence, Italy  \\
\hspace*{1mm}
$^{ 3}$ now at Univ. of Salerno and INFN Napoli, Italy  \\
\hspace*{1mm}
% ...
$^{12}$ now with OPAL Collaboration, Faculty of Physics at Univ. of
        Freiburg \\
% ...

\begin{tabular}{lp{15cm}}
$^{a}$ &supported by the Natural Sciences and Engineering Research
         Council of Canada (NSERC) \\
$^{b}$ &supported by the FCAR of Quebec, Canada\\
$^{c}$ &supported by the German Federal Ministry for Research and
         Technology (BMFT)\\
$^{d}$ &supported by the MINERVA Gesellschaft f\"ur Forschung GmbH,
         and by the Israel Academy of Science \\
$^{e}$ &supported by the German Israeli Foundation, and
         by the Israel Academy of Science \\
$^{f}$ &supported by the Italian National Institute for Nuclear Physics
         (INFN) \\
$^{g}$ &supported by the Japanese Ministry of Education, Science and
         Culture (the Monbusho)
         and its grants for Scientific Research\\
$^{h}$ &supported by the Korean Ministry of Education and Korea Science
         and Engineering Foundation \\
$^{i}$ &supported by the Netherlands Foundation for Research on Matter
         (FOM)\\
$^{j}$ &supported by the Polish State Committee for Scientific Research
         (grant No. SPB/P3/202/93) and the Foundation for Polish-
         German Collaboration (proj. No. 506/92) \\
$^{k}$ &supported by the Polish State Committee for Scientific
         Research (grant No. PB 861/2/91 and No. 2 2372 9102,
         grant No. PB 2 2376 9102 and No. PB 2 0092 9101) \\
$^{l}$ &partially supported by the German Federal Ministry for
         Research and Technology (BMFT) \\
$^{m}$ &supported by the German Federal Ministry for Research and
         Technology (BMFT), the Volkswagen Foundation, and the Deutsche
         Forschungsgemeinschaft \\
$^{n}$ &supported by the Spanish Ministry of Education and Science
         through funds provided by CICYT \\
$^{o}$ &supported by the Particle Physics and Astronomy Research
        Council \\
$^{p}$ &supported by the US Department of Energy \\
$^{q}$ &supported by the US National Science Foundation
\end{tabular}

\newpage
\pagenumbering{arabic}
\setcounter{page}{1}
\normalsize

\section{Introduction}
\label{s:intro}
A widely studied topic of perturbative Quantum
Chromodynamics (pQCD)
is the nature of the quark fragmentation
process from the quark-antiquark pair created in \ee~annihilation experiments.
The purpose of this paper is to compare the fragmentation properties of
the struck quark in Deep Inelastic Scattering (DIS) to those of the
quarks produced in such \ee~annihilation experiments, in order to test the
universality of the quark fragmentation process.

Operation of the HERA electron-proton collider has vastly
extended the kinematic range for studies of the hadronic final state in
DIS. The event kinematics of DIS
are determined by the negative square of the four-momentum transfer,
$Q^2\equiv-q^2$, and the Bjorken scaling variable, $x=Q^2/2P\!\cdot\!q$,
where $P$ is the four-momentum of the proton.
In the Quark Parton Model (QPM),
the interacting quark from the proton carries the four-momentum $xP$.
The variable $y$, the fractional energy transfer to the proton in its rest
frame, is related to $x$ and $Q^2$ by $y=Q^2/xs$, where $\sqrt s$ is
the electron-proton centre of mass energy.

A natural frame to study the dynamics of the hadronic final state
in DIS is the Breit frame~\cite{feyn}.\linebreak
In this frame the exchanged
virtual boson is completely space-like and is given by \linebreak
\mbox{$q = (0,0,0,-Q=-2xP^B)\equiv (E,p_x,p_y,p_z)$}
where $P^B$ is the proton momentum in the Breit frame.
In the QPM the $z$-component of the momentum of the incoming
quark is $Q/2$ before and $-Q/2$ after the interaction
with the exchanged virtual boson.
The particles produced in the DIS interaction can then be assigned to one of
two regions. The current region corresponds to
the direction of the outgoing struck quark in the QPM.
In this paper we define the current region corresponding to
particles with $p_z<0$, whilst the target region is defined by $p_z>0$.

In \ee~annihilation the two quarks are produced
with equal and opposite momenta, $\pm \sqrt{s}/2.$
In the Breit frame of DIS, a quark is struck from within the
proton with outgoing momentum $-Q/2$.
In the direction of the struck quark
the particle spectra are expected to have no
dependence on $x$ and a
dependence on $Q$ similar to those observed in
\ee~annihilation~\cite{eedis,anis,char} at energy $\sqrt{s}=Q.$

The results from \ee~annihilation support the inclusion of coherence
effects in pQCD~\cite{bassetto,mueller,webber,MLLA}.
The phenomenon of coherence is a natural consequence of the quantum mechanical
treatment of the parton cascade.
As long wavelength gluons are unable to resolve individual colour charges
of partons within the parton cascade, the available phase space for soft gluon
emissions is reduced to an angular-ordered region, due to destructive
interference. This leads to a number of important differences in the
properties of the partonic final state relative to the incoherent case.
The most notable of these are the slower rise in the multiplicity of
partons with increasing energy and the modification of the logarithmic
parton momentum spectra to an approximately
Gaussian form, which is often referred to as the ``hump-backed''
plateau~\cite{MLLA}.

Coherence effects are explicitly included in the Modified
Leading Log Approximation (MLLA)~\cite{eedis} of pQCD.
The MLLA calculations predict the parton
multiplicity and the form of the momentum spectra from quark
fragmentation at a given, sufficiently large, energy.
The hypothesis of
Local Parton Hadron Duality (LPHD)~\cite{LPHD},
which relates the observed hadron
distributions to the calculated parton distributions via a constant of
proportionality $\kappa^{\rm ch}$, is used in conjunction with the predictions
of the MLLA. Using this assumption the modifications due to coherence
are therefore not only expected to be experimentally observed, but the MLLA
calculations should also be directly applicable to data.

So far these effects have been studied at \ee~annihilation experiments
\cite{tasso,opal,L3} using the
average charged particle multiplicity, $<\!n_{ch}\!>$, and the
distribution of the scaled momentum,
\logxp, where $x_p=2p/\sqrt{s}$.
The results from these experiments are in accordance with the MLLA
and the assumptions of LPHD.

Previous fixed-target DIS ($\bar \nu p$) measurements in the Breit
frame
have been
performed in the range
\mbox{$1<Q^2<45~$GeV$^2$} with a mean $\bar \nu$ beam energy
of $33$ GeV~\cite{nubarp}.
In the present paper the scaled momentum and charged multiplicity
distributions of the hadronic final state are
measured in the current region of the Breit frame
as a function of $x$ and $Q$ in the range
$6\times10^{-4} < x < 5\times10^{-2}$ and \mbox{$10<Q^2<1280~$GeV$^2$}
with a mean energy of the electron beam measured in the proton
rest frame $\sim 10^4$~GeV.
Comparisons are made with Monte Carlo
models, MLLA analytic calculations and \ee~data.
The DIS data were obtained in 1993 with the ZEUS detector at the HERA collider
where electrons of energy $E_e=26.7~{\rm GeV}$ collided with protons of
energy $E_p=820~{\rm GeV}$, and correspond to an integrated luminosity of
0.55~pb$^{-1}$.

\section{The ZEUS detector and trigger}

ZEUS is a multipurpose magnetic detector which
has been described
elsewhere \cite{zeus2,zeus3}. Here
we give a brief description concentrating on
those parts of the detector relevant for the present analysis.

Charged particles are
tracked by the inner tracking detectors which operate in a
magnetic field of 1.43 T provided by a thin superconducting coil.
Immediately surrounding the beampipe is the vertex detector (VXD) which
consists of 120 radial cells, each with 12 sense wires~\cite{vxd}.
The achieved resolution is $50~\mu$m in the central region of a cell
and $150~\mu$m near the edges.
Surrounding the VXD is
the central tracking detector (CTD) which consists of 72 cylindrical
drift chamber layers,  organised into 9 `superlayers'~\cite{ctd}.
These superlayers alternate between
those with wires parallel (axial) to the collision axis
and those inclined at a small angle to give a stereo view.
The magnetic field
is significantly inhomogeneous towards the ends of the CTD thus
complicating the electron drift.
With the present understanding of the chamber, a spatial resolution of
$260~\mu$m has been achieved.
The hit efficiency of the chamber is greater than 95\%.

In events with charged
particle tracks, using the combined data from both chambers,
reconstructed primary vertex position resolutions
of $0.6$ cm in the $Z$ direction
and $0.1$ cm in the $XY$ plane are measured.
(The ZEUS coordinate system
is defined as right handed with the $Z$ axis pointing in the
proton beam direction and the $X$
axis horizontal pointing towards the centre of HERA.
The polar angle $\theta$ is defined with respect to the $Z$-direction.)
The resolution in transverse momentum
for full length tracks is $\sigma_{p_T}/p_T=0.005p_T\oplus 0.016$
(for $p_T$ in GeV).

The solenoid is surrounded by a high resolution  uranium-scintillator
calorimeter divided into three parts,
forward (FCAL),
barrel (BCAL) and rear (RCAL).
Holes of $20\times 20$ cm$^2$ in the centre of FCAL and
RCAL are required to accommodate the HERA beam pipe.
Each of the calorimeter parts is
subdivided into towers which in turn are segmented longitudinally
into electromagnetic
(EMC) and hadronic (HAC) sections. These sections are further
subdivided into cells,
which are read out by two photomultiplier tubes.
A detailed description of the calorimeter is given in \cite{test1}.

For measuring the luminosity as well as for tagging very small $Q^2$
processes, two lead-scintillator calorimeters are used~\cite{lumi}.
Bremsstrahlung photons emerging from the electron-proton interaction
point (IP) at angles $\theta_\gamma \le 0.5$ mrad with respect to the
electron beam axis hit the photon calorimeter at 107 m from the IP.
Electrons emitted from the IP at scattering angles
less than 6~mrad and with energies, $E_e^{\prime}$, between $0.2E_e$ and
$0.9E_e$ are deflected by beam magnets and
hit the electron calorimeter placed 35~m from the IP.

For events with the scattered electron detected in the calorimeter, the
trigger was essentially independent of the DIS hadronic final state. The
trigger acceptance was greater than 97\% for $Q^2 > 10~\mbox{\rm GeV}^2 $
and independent of $Q^2$~\cite{FLT}.

\section{Event selection}
\label{s:data_proc}

The offline selection of DIS events was similar to that described in
our earlier publications \linebreak[4]
[21-26].
% \cite{f2,zeus:efl,1992f2,zeus:rapgap,jetgap,rap_e_flow}.
Scattered electron candidates
were selected by using the pattern of energy deposition in the
calorimeter.
The electron identification algorithm was
tuned for purity rather than efficiency.
In studies with Monte Carlo DIS events and test beam data the
purity was estimated to be $\geq 96~\%$ for $E_e^\prime \ge 10~{\rm GeV}$.

The ZEUS detector is almost hermetic, allowing the kinematic
variables $x$ and $Q^2$ to be reconstructed in a variety of
ways using combinations of electron and hadronic system energies and
angles. Measurements made purely from the measurement of the
scattered electron energy and angle, are denoted with a subscript $e$
whilst those made purely from the hadronic system, by the
Jacquet-Blondel method, are denoted by the subscript $JB$~\cite{jb}.
The double angle
method which calculates the kinematic variables from the scattered electron
angle and the angle of the struck quark, $\gamma_H$, as calculated
from the final hadronic system are denoted by the subscript $DA$ \cite{DA}.

In the kinematic region selected, the optimal method of determining $Q^2$, $x$
and the boost, $\vec{\beta}$, from the laboratory to the Breit frame is based
on the $DA$ method.

For our final event selection we demanded
\begin{itemize}
\item
$E^\prime_e \ge 10~{\rm GeV}$,
to achieve a high purity sample of DIS events;
\item
$Q^2_{DA}\geq 10$ GeV$^2$;
\item
$y_e\leq 0.95$,
to reduce the photoproduction background;
\item
$y_{JB}\geq 0.04$, to give sufficient accuracy for $DA$ reconstruction;
\item $\delta = \sum_i E_i(1 - \cos\theta_i)  \ge 35$~GeV,
       where the sum runs over all
      calorimeter cells.
      The measured energy of the cell is denoted by
      $E_i$ and its polar angle with respect to the incident proton beam
      by $\theta_i$.
      Nominally, $\delta$ should peak at twice the electron beam energy,
      $53.4$~GeV.
      This cut is used to remove photoproduction events and
      to control radiative corrections.
\end{itemize}
%page break by hand
\newpage
Furthermore we required
\begin{itemize}
\item
a primary vertex position, determined from VXD and CTD tracks, in the
range \linebreak
$-50 \leq Z_{vtx} \leq \rm 40~cm$ and a cut on the radial distance from
the beamline \linebreak
$R_{vtx}~= \sqrt{X^2_{vtx} + Y^2_{vtx}}<~10~\rm cm$;
\item
the impact point $(X,Y)$ of the scattered electron in the RCAL to lie
outside a square of $32 \times 32$ cm$^2$ centred on the beam axis, to
ensure the electron is fully contained within the detector and its position can
be reconstructed with sufficient accuracy;
\item
no more than 5~GeV of energy deposition in the electron calorimeter of the
luminosity detector, to exclude potential photoproduction background events.
\end{itemize}
Following these cuts, the remaining photoproduction background was
estimated to be $\simeq$1\% using events tagged in the luminosity detector.
The contamination from beam-gas background was estimated to be below 0.5\%
from unpaired electron and proton bunches.
Finally, we rejected QED Compton scattering events and residual cosmic
and beam-related muons.

A total of 31.5K events was selected in this way corresponding to an
integrated luminosity of 0.55~pb$^{-1}$.
Of these events about 8-10\% \cite{zeus:rapgap}
contain a large rapidity gap
in the hadronic final state.
These are characterised by
$\eta_{max}<1.5$, where $\eta_{max}$ is the maximum pseudorapidity of
any calorimeter cluster in the event.
Here, the
pseudorapidity is defined by $\eta = -{\rm ln(tan(\theta/2))}$ and
a cluster is an isolated set
of adjacent cells with summed energy above 400 MeV.

\section{Track reconstruction and selection}
\label{s:track_proc}

Two programs for track finding and fitting have been developed independently
which follow different strategies in the pattern recognition and track fitting
stages.

In the first approach, adopted for the final data analysis,
the track finding algorithm starts with
hits in the  outermost axial superlayers of the CTD.  As the
trajectory is followed inwards to the beam axis,
more hits from the axial wires and the VXD are incorporated.
The resulting circle in the $XY$ projection is used for the pattern
recognition in the stereo superlayer pattern.
The momentum vector is determined in a 5-parameter helix fit.

The other track finding program
is based on the Kalman filtering technique~\cite{Kalman}. Seed tracks found
in the outer
layers of the CTD are extended  inwards and points are added as wire
layers of the CTD are crossed. The track
parameters at each step are updated using the Kalman method.
In the second step a Kalman fit to the points found in the pattern
recognition phase is performed taking into account non-linear corrections
to the measured drift time.
Following the reconstructed CTD track inwards, CTD and VXD hits
are associated to the track.
The CTD tracks are merged to VXD track segments
using the Kalman filtering algorithm.

Multiple Coulomb scattering in the beampipe,
and the walls of the VXD and CTD
were taken into account in the evaluation of the covariance matrix.
The vertex fit is performed with the fitted tracks using the
perigee parameterisation \cite{billoir}. The vertex position is evaluated and
the track parameters at the vertex are re-evaluated.

The reconstructed tracks used in this analysis
are associated with the primary event vertex
and have $p_T>200$~MeV and $|\eta|<1.5$, corresponding to
the polar angle region between $25^{\circ}$ and $155^{\circ}$.
This is a region of good CTD acceptance
where the detector response and systematics are best understood.
For tracks within the fiducial volume defined by these cuts, the track
reconstruction efficiency is $\simeq$~95\%.

The Breit frame boost was reconstructed using the scattered electron, where
the energy was determined using the $DA$ formula and the
polar and azimuthal angles were measured from the impact point
on the calorimeter.
The four-momentum vectors of the charged particles were boosted to the Breit
frame, assuming the pion mass to determine the particle's energy, and
were assigned to the current region if $p_z<0$ in the Breit frame.

Monte Carlo simulations were used to determine
the acceptance for tracks in the current region as a function of $(x,Q^2)$.
The chosen analysis intervals in $(x,Q^2)$ correspond to regions of
high acceptance $\sgeq 70\%$ in the current region of the Breit frame.
These intervals are commensurate with the resolution of $x$ and
$Q^2$.

The measured variables in the analysis are the
inclusive mean charged multiplicity, $<\!n_{ch}\!>$,
in the current region of the Breit frame
and the distributions of \logxp.
Here, $x_p = 2p/Q$, which is the momentum $p$
of a track from the primary vertex measured in the Breit frame,
scaled by $Q/2$, the maximum possible momentum
(ignoring effects due to intrinsic $k_T$ of the quark within the
proton). The peak position of the \logxp~distribution is denoted by \logxpmax.

Uncertainty in the reconstruction of the boost vector, $\vec{\beta}$,
was found to be the most significant factor
on the resolution of \logxp.
Near the peak position of the \logxp~distributions the resolution
is $\simeq 0.15$ units, with no significant shift around the peak position,
leading to a choice of bin width of 0.25 units.
Migration of tracks from the current region to the target region
was typically $\simeq 8\%$. The migration into the current region from
the target fragmentation region is of a similar magnitude as the migration
out. At low values of $y$, however, the level of migration
is up to $\simeq 25\%$.  In the low $y$ region, the hadronic activity is low
and the measurement of $\gamma_H$ becomes distorted by noise in the
calorimeter leading to a reduced $x$ resolution
and hence an uncertainty in $\vec{\beta}$.

\section{QCD models and event simulation}
\label{s:model}
Monte Carlo event simulation is used to correct for acceptance and
resolution effects.  The detector simulation is based on the
GEANT~3.13~\cite{GEANT} program and incorporates our best knowledge of the
apparatus. Details of the
detector (MOZART) and trigger (ZGANA) simulation codes are given in
\cite{zeus2}.

Neutral current DIS events with $Q^{2}>4~\Gevsq$
were generated using the HERACLES~4.4 \mbox{program}~\cite{HERACLES} which
incorporates first order electroweak corrections.
The Monte Carlo generator LEPTO~6.1 \cite{LEPTO},
interfaced to HERACLES via the program DJANGO~6.0~\cite{DJANGO},
was used to simulate QCD cascades and fragmentation.
The parton cascade was modelled
in different ways:
\begin{itemize}
\item{
with the colour-dipole model including the boson-gluon fusion process (CDMBGF),
using the ARIADNE~4.03 \cite{ariadne} program. In this model coherence
effects are implicitly included in the formalism of the parton cascade; and,}
\item{using the matrix element plus parton showers option (MEPS) within
LEPTO,
where coherence effects in the final state cascade are included by angular
ordering of successive parton emissions.}
\end{itemize}

These models use the Lund string fragmentation model \cite{string}
for the hadronisation phase as implemented
in JETSET~7.3~\cite{JETSET}.
An additional sample of events was generated with the HERWIG~5.7 Monte Carlo
\cite{herwig},
%with the soft underlying event option,
where no electroweak radiative
corrections were applied. The parton cascade includes coherence effects
in a manner similar to that of LEPTO
and a clustering model is used for the hadronisation \cite{webber,cluster}.

For ARIADNE and HERWIG the parameterisation of the parton distribution
functions was the $\rm{MRSD_-^{\prime}}$ set~\cite{mrsdm}.
The GRV~\cite{grv} parameterisation was used for the MEPS data set.
These parameterisations
have been shown to describe reasonably the HERA measurements
of the proton structure function, $F_2$ \cite{f2,h1f2}.

Whilst these programs give a reasonable description
of the observed energy flow \cite{zeus:efl} they do not describe the
excess of events observed with a large rapidity gap.
The properties of these diffractive events are consistent
with the exchange of a pomeron between the proton and virtual
photon. The POMPYT Monte Carlo \cite{pompyt} models high-energy
diffractive processes where
the proton emits such a pomeron whose constituents take part in a
hard scattering process with the virtual photon. A hard quark density
distribution for the pomeron provides an acceptable description of this class
of events \cite{rap_e_flow}. The Monte Carlo program based on the model
of Nikolaev-Zakharov \cite{NZ} also gives an acceptable
description of this class of events.

For each of the above Monte Carlo programs the default parameters were used.
The Monte Carlo event samples were passed through reconstruction and
selection procedures identical to those for the data.

\section{Data correction}
\label{s:correction}
The correction procedure is based on the detailed Monte Carlo
simulation of the ZEUS detector with the event generators
described in the previous section.
Since the ARIADNE model gives the best overall description of our observed
energy flow~\cite{zeus:efl} it is used for the standard corrections and
unfolding of the distributions.

The data are corrected for trigger and event selection cuts;
event migration between ($x,Q^2$) intervals;
QED radiative effects;
track reconstruction efficiency;
track selection cuts in $p_T$ and $\eta$;
track migration between the current and target regions;
and for the decay
products of $K^{0}_S$ and $\Lambda$ decays which are assigned to the
primary vertex.

Correction factors were obtained from the Monte Carlo simulation by
comparing the ``true'' generated distributions before the detector
and trigger simulations
with the ``observed'' distributions after these simulations followed
by the same reconstruction, selection and analysis
as the real data.
The ``true'' distributions did not include the charged particle decay
products of $K^0_S$ and $\Lambda$ and
charged particles produced from weakly decaying particles with a
lifetime \mbox{$>10^{-8}$s}. Monte Carlo studies showed that up to
2\% of the current region tracks from the reconstructed
primary vertex are due to charged particles from the
decay of $K^0_S$ and $\Lambda$.
In a separate analysis, the shape of the $K^0_S$ and $\Lambda$
distributions are shown to be well reproduced by
the Monte Carlo \cite{k0}.

For the \logxp~distributions,
correction factors were calculated for each \logxp~bin
$$
F(\logxp\!) = \frac{1}{N_{\rm gen}} \; \left( \frac{dn}{d\ln(1/x_p)} \right)
_{\rm gen}
 \left/ \frac{1}{N_{\rm obs}} \; \left( \frac{dn}{d\ln(1/x_p)} \right) _{\rm
obs}  \right.
$$
where $N_{\rm gen}$ ($N_{\rm obs}$) is the number of generated (observed)
Monte Carlo events in the $(x,Q^2)$ interval.
The overall correction factors are greater than unity and
typically $<1.3$ and are independent of
\logxp~around the peak position.

The correction procedure for the multiplicity distributions
was performed in two stages. The dependence of this procedure on the
Monte Carlo input is discussed in section~\ref{s:systematics}.
The first stage was to correct for track reconstruction efficiency
and track migrations between the current and target regions
in each $(x_{rec},Q^2_{rec})$ interval where
$x_{rec}$ and $Q^2_{rec}$
are the reconstructed values of $x$ and $Q^2$.
In each interval a comparison was made between the
observed multiplicity distribution
$P_{n_{\rm o}}(x_{rec},Q^2_{rec})$, and the generated distribution
$P_{n_{\rm p}}(x_{rec},Q^2_{rec})$.
This comparison yielded the
correction matrix $M_{n_{\rm p}, n_{\rm o}}(x_{rec},Q^2_{rec})$
with elements defined by
$$
M_{n_{\rm p},n_{\rm o}}(x_{rec},Q^2_{rec})=\frac{{\rm
No.\;of\;events\;with\;}n_{\rm p}\;{\rm tracks\;generated\;when}\;n_{\rm
o}\;{\rm tracks\;were\;observed}}{{\rm No.\;of\;events\;with} \;n_{\rm o}\;{\rm
tracks\;observed}}\, .
$$
This matrix relates the observed to the generated distributions in each
$(x_{rec},Q^2_{rec})$ interval by
$$
P_{n_{\rm p}}(x_{rec},Q^2_{rec}) = \sum_{n_{\rm o}}
M_{n_{\rm p},n_{\rm o}}(x_{rec},Q^2_{rec})
\cdot P_{n_{\rm o}}(x_{rec},Q^2_{rec})\, .
$$

The second stage corrected for migrations between the
analysis intervals and for the acceptance of the event selection cuts
using the correction factors
$$
C=\frac{\rho_{true}(x_{true},Q^2_{true})}
{\rho_{rec}(x_{rec},Q^2_{rec})}
$$
where the $\rho$'s are normalised multiplicity distributions in each
interval at the reconstructed level ($rec$) and at the
generator level where no selection cuts are applied ($true$).

The corrected multiplicity distribution was then calculated
according to the formula
$$
P_{n_{\rm p}}(x_{true},Q^2_{true})= C \cdot
\sum_{n_{\rm o}}
M_{n_{\rm p},n_{\rm o}}(x_{rec},Q^2_{rec})\cdot P_{n_{\rm
o}}(x_{rec},Q^2_{rec})
$$

The overall correction factors are in the range $1.25-1.5$ and are dominated
by the corrections from applying the
matrix $M_{n_{\rm p}, n_{\rm o}}(x_{rec},Q^2_{rec})$.
In the lowest two $Q^2$ intervals the
corrections from $C$ are of the same
size as those from the matrix.
The corrected multiplicity distributions in each $(x,Q^2)$ interval are
henceforth denoted by $P(n_{ch})$.

\section{Systematic errors}
\label{s:systematics}
A number of systematic checks were performed in order to investigate the
sensitivity of the corrected results to features of the analysis as
described previously.
These can be categorised under event selection, track and vertex
reconstruction, track efficiency and selection, and data correction methods.

%Event selection
%===============
Systematic uncertainties arising from remaining photoproduction
background were studied by tightening the $y_{e}$ cut from 0.95 to 0.8 and the
$\delta$ cut from 35~GeV to 40~GeV.
The effects of calorimeter noise on the reconstruction of $x$
were checked by tightening the $y_{JB}$ cut
from 0.04 to 0.05. The effect of a small mismatch in electron energy scale
between the Monte Carlo and data was checked by scaling the energy of the
calorimeter cells associated with the electron in the Monte Carlo.
The overall systematic errors associated with the event selection
were negligible.

The analysis was repeated using the second approach to pattern recognition,
track and vertex fitting (see section~\ref{s:track_proc}) applied to both the
Monte Carlo sample and the data. This was found to be a significant source of
systematic error.
The mean multiplicity results changed by $3-5\%$ in general but the effect was
up
to $8\%$. Changes to \logxpmax~were typically $3\%$ although
the maximum shift observed was $9\%$.

The efficiency of the tracking system was varied within the Monte Carlo by
removing all hits from the vertex detector and $80\%$ of hits from a random
superlayer in the CTD. This is  a conservative
estimate of the systematic uncertainty due to inefficiencies within the
detector.
The largest change to the mean multiplicity was $3-5\%.$
Removing the $p_T$ and $\eta$ track selection cuts produced
changes in the results which were within the statistical errors.
(The intervals in $(x,Q^2)$ were chosen in order to maximise the
acceptance in the current region and therefore reduce the
sensitivity to these cuts.)
The changes due to varying the parameters in the vertex fit
were again within statistical errors.

The Monte Carlo-based correction techniques described in section
\ref{s:correction} are model dependent. In order to
investigate the size of this effect the data were corrected with two
independent Monte Carlo samples, MEPS and HERWIG. The differences
in the simulation of the hadronic final state
implicitly test the uncertainty due to the boost reconstruction
which is derived from the $DA$ method.
The systematic uncertainties on
\logxpmax~and the mean multiplicity were typically between $4-7\%$.
This is  the
largest single contribution to the overall systematic error.
In the highest $(x,Q^2)$ interval,
where the statistics from Monte Carlo and data are limited,
the largest change to the mean multiplicity was $17\%$.

Inclusion of events with a large rapidity gap in the correction procedure
was investigated by correcting the data with a sum of
$90\%$ ARIADNE and $10\%$ POMPYT events which reasonably describes the DIS
data~\cite{jetgap}.
Such events are a small
contribution to the inclusive distribution and as a result the \logxpmax~values
are reduced by $1\%$ whereas the mean multiplicities are more
sensitive and increase by $3-5\%$ due to the effects discussed in
section~\ref{s:multi}.

The systematic deviations from the
above checks were combined in quadrature to yield the quoted
systematic errors.
\section{Results}
\subsection{Multiplicity}
\label{s:multi}
\begin{figure}[tp]
\centerline{\psfig{figure=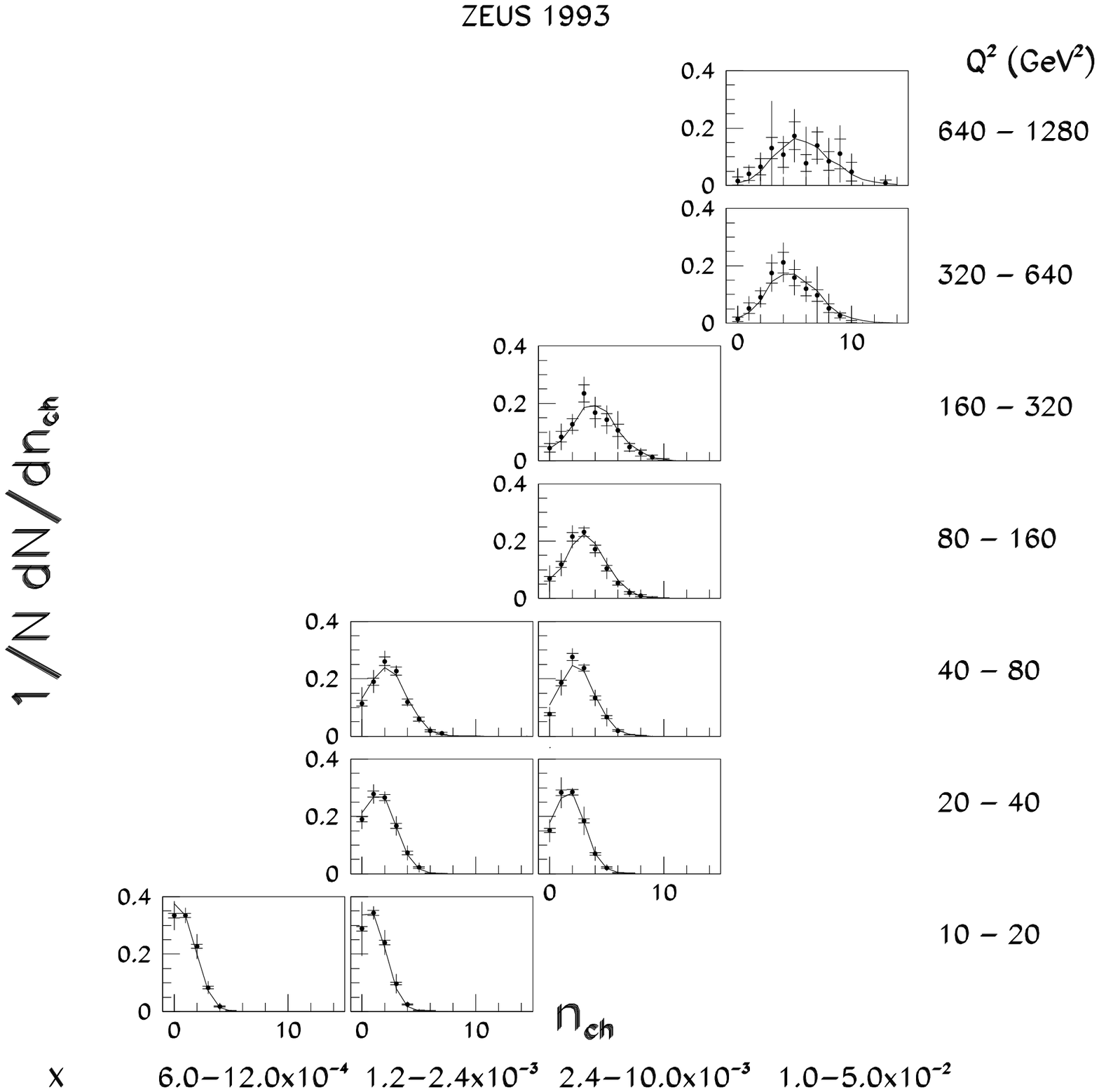,width=18cm}}
\caption{{\it Charged multiplicity distributions
in the current region of the Breit frame as a function of $(x,Q^{2})$.
Statistical errors are indicated by the inner error bar bounded by
the horizontal bars. The outer error bars show the statistical and
systematic errors added in quadrature.
The ZEUS data are compared to the predictions of the ARIADNE Monte
Carlo.}}
\label{figure:over_cdm}
\end{figure}

The variables and distributions presented in the following sections have
been corrected for detector and acceptance effects.
The multiplicity distributions are
binned in $x$ and $Q^2$ as indicated in table~\ref{table:meannch},
in the intervals
$6\times10^{-4} < x < 5\times10^{-2}$ and \mbox{$10<Q^2<1280~$GeV$^2$},
where the mean values of $Q$ range from $\simeq 4$ to 30~GeV.
Figure~\ref{figure:over_cdm} shows the measured multiplicity distributions
in the current region of the Breit frame.
With increasing $Q$ the mean of the multiplicity increases.
At low $Q$, there is a significant probability for no tracks to be found in
the current region which we address later.
The multiplicity distributions are compared to the ARIADNE Monte Carlo which
reproduces the data in each $(x,Q^2)$ interval.

\begin{figure}[tp]
\centerline{\psfig{figure=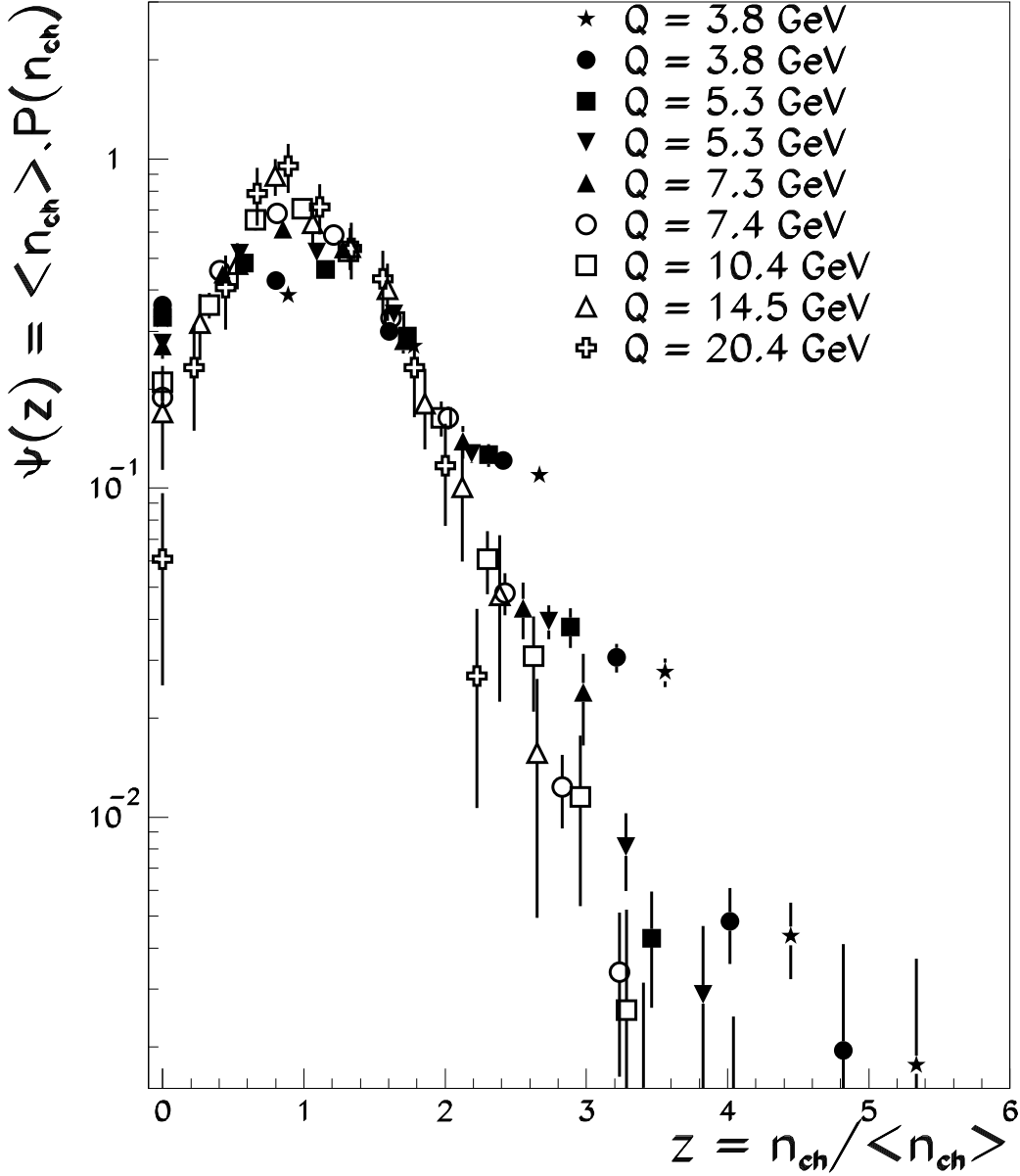,width=18cm}}
\caption{{\it The KNO distribution in the current region of the Breit frame
as measured by ZEUS.
The values of $Q$ correspond to the $x$-intervals shown in table~1.
The highest $(x,Q^2)$ interval is omitted for the sake of clarity.
Statistical errors only are indicated.}}
\label{figure:kno}
\end{figure}

In order to compare the multiplicity distributions at various values of $Q$,
the scaled multiplicity distributions, $\Psi(z) = <\!n_{ch}\!>\cdot P(n_{ch})$,
are plotted as a function of $z = {n_{ch}\over{<\!n_{ch}\!>}}$
in figure~\ref{figure:kno}~\cite{KNOREF}.
For $Q \sleq 7$~GeV, the distributions do not scale with $Q$.
For $Q \sgeq 7$~GeV, the scaling violations are less prononunced,
the distributions becoming
narrower with increasing $Q$ indicating scaling violations of the KNO
variable~\cite{cmulti} similar to those observed in
\ee~annihilation~\cite{nch:tas}.

The mean multiplicities, $<\!n_{ch}\!>$, are shown in table~\ref{table:meannch}
as
a function of $(x, Q^2)$. Over the measured range of $Q$,
the mean multiplicity increases by a factor of five.
Comparison of the mean multiplicity in
intervals of fixed $Q$ and differing $x$ provides a check on a possible
$x$ dependence in the current region of the Breit frame. The mean
multiplicities
in these intervals are the same within systematic errors.

\begin{table}[htb]
\begin{center}
\begin{tabular}{c r@{-}l r@{.}l c  l} \hline
$x$ range                & \multicolumn{2}{c}{$Q^2$ range}
&\multicolumn{2}{c}{$Q$ mean}& $<\!n_{ch}\!>\pm$ stat $\pm$ sys \\
                         & \multicolumn{2}{c}{(GeV$^2$)} &
\multicolumn{2}{c}{(GeV)}   &                             \\ \hline \hline
$0.6- 1.2\times 10^{-3}$  &  10&20                  & 3&8
& $1.12\pm 0.02\pm 0.15$   \\ \\
$1.2- 2.4\times 10^{-3}$  &  10&20                  & 3&8
& $1.25\pm 0.02\pm 0.25$   \\
                          &  20&40                  & 5&3
& $1.73\pm 0.04\pm 0.23$   \\
                          &  40&80                  & 7&3
& $2.35\pm 0.07\pm 0.11$   \\ \\
$0.24-1.0\times 10^{-2}$  &  20&40                  & 5&3
& $1.83\pm 0.03\pm 0.27$   \\
                          &  40&80                  & 7&4
& $2.47\pm 0.05\pm 0.23$   \\
                          & 80&160                  & 10&4
& $3.05\pm 0.09\pm 0.31$   \\
                          & 160&320                  & 14&5
& $3.77\pm 0.22\pm 0.40$   \\ \\
$1.0- 5.0\times 10^{-2}$  & 320&640                  & 20&4
& $4.49\pm 0.30\pm 0.54$   \\
                          & 640&1280                  & 29&2
& $5.57\pm 0.65\pm 1.21$   \\ \hline
\end{tabular}
\end{center}
\caption{\label{table:meannch}{\it Mean charged multiplicity
in the current fragmentation region.}}
\end{table}

In figure~3a the mean charged multiplicity as a function of $Q$
is compared to the predictions of three Monte Carlo models
which incorporate coherence effects. The MEPS and ARIADNE models
reproduce the growth in multiplicity as observed in the data.
These Monte Carlo generators  predict a higher multiplicity for
the higher $x$ range at fixed $Q$ values.
HERWIG, which includes coherent parton showers but has no explicit matrix
element, also reproduces the increase with $Q$.
Coherence suppresses the production of large multiplicity events
which becomes a more noticeable effect with increasing $Q$.

The effects of coherence on the evolution of the parton shower have been
tested by calculating the mean multiplicity with the MEPS model
(with default parameter settings),
as shown in figure~3b.
The MEPS incoherent parton shower with
independent fragmentation exhibits a faster
growth of the multiplicity than seen in the data.
The discontinuities in the lines correspond to predictions for
two different $x$ values at fixed $Q$.
The growth of the
incoherent case is damped when string fragmentation is used rather than
independent fragmentation but not enough to agree with the data.
Additional low-momentum partons produced in
the incoherent shower correspond to small kinks of the Lund string.
These therefore lead to relatively small differences between
the incoherent and coherent case when string fragmentation is
applied~\cite{boudinov}.
Overall, however, the data are best described by Monte Carlo models that
include coherence in their simulations, whilst those with an
incoherent shower are less successful in reproducing
the multiplicity distributions and the growth of $<\!n_{ch}\!>$ with $Q$.

The definition used here for associating particles with the current
fragmentation region, namely $p_z < 0$, should be considered an operational
definition which is unambiguous only when the produced jet and the
fragmentation
products are massless. If these are massive and/or if QCD radiation
is included, the current and target fragmentation regions begin to merge.
Consider for example the case where the struck quark branches into a quark and
a gluon producing a ($qg$) system with mass $m$. Being massive, the
$z$-momentum of the $qg$ system becomes more positive; a special
case is $m=Q$ where the $qg$ system is at rest in the Breit frame.
If, instead of the definition used in this paper, one would take the
point of view that the current fragmentation should include the
particles emerging from the hadronisation of the outgoing quark $and$
gluon then the current region defined by $p_z < 0$ would contain only
part of the fragmentation products.
The Bremsstrahlung nature of QCD radiation results in
partons which are mostly soft and collinear leading to small masses~$m$.
Clearly, the effect should be small if $Q>>m$.
Two exercises were made in order to estimate the
size of the effect, keeping the definition of the current region as
before, namely $p_z < 0$.

In the first study, the multiplicities in the current region predicted
by the QPM and the ME model (i.e. first-order QCD matrix elements without
parton showers)  were compared. The fragmentation was performed in both
cases according to the Lund scheme. In the ME calculation, QCD branching
is considered if $y_{ij}~>~y_{cut}$, where $y_{ij}~=~m_{ij}^2/W^2$.
Here $m_{ij}$ is
the mass of the parton system produced in the branching and $W$ is the total
centre of mass energy available for hadron production. The calculation
was performed choosing a standard value of $y_{cut}~=~0.015$. Note that
this choice of $y_{cut}$ corresponds to a hard branching which is a
relatively rare process. The average multiplicities predicted by ME at
$Q=3.8$~GeV were found to be $\simeq$12\% lower than those of QPM, the
difference
becoming smaller as $Q$ increased.

In the second study, this effect was investigated with data by comparing
events with and without a large rapidity gap. For large rapidity gap
(LRG) events, QCD radiation is suppressed compared to events without a
large rapidity gap~\cite{rap_e_flow}. Therefore, the depopulation of
the current region due to QCD radiation should also be suppressed in
LRG events. Indeed, in the lowest $Q$ interval (\mbox{$Q=3.8$~GeV})
LRG events were observed to have an $\simeq 50\%$ higher multiplicity
compared with non-LRG; for \mbox{$Q=7.3$~GeV} the excess was
$\simeq 15\%$ and for $Q~>~10$~GeV the same multiplicity was
observed for both event classes, within errors.

The preceding considerations suggest that for $Q~>~10$~GeV the different
definitions of the current region lead to approximately the same results
for current fragmentation and that the data obtained here from DIS can be
directly compared with those on quark fragmentation measured in
\ee~annihilation.

In figure~\ref{figure:nchepem}, twice the mean
multiplicity is compared with inclusive mean multiplicity
measurements from \ee~experiments
\cite{nch:tas,nch:plu,nch:opl,nch:hrs} for $Q \equiv \sqrt{s} > 10$~GeV.
In the region of overlap
the results from ZEUS and the \ee~experiments are
in agreement, exhibiting a similar rate of growth with $Q$.
DIS provides a larger fraction of light quarks
compared with the \ee~case~\cite{anis}.
The different flavour composition of the
two physics processes could, in principle, affect the comparison of
these measurements.
However, Monte Carlo generator studies of heavy quark
production in \ee~using JETSET
and results from \ee~\cite{opal_hvy} on  $<\!n_{ch}\!>$ from
a sample excluding $b$-quarks
show that these effects give rise to
differences in the results that are within the quoted ZEUS
errors.

\subsection{Scaled momentum spectra}
\label{s:logxp}
In figure~\ref{figure:corr_data} the
\logxp~distributions for charged particles are shown
in the same $(x,Q^2)$ intervals as the multiplicity distributions.
These distributions are approximately Gaussian in shape with
mean multiplicities given by the integral of the distribution (this estimate
agrees within errors with the multiplicity results, but differs slightly
due to differences in the correction methods).
The \logxp~distributions are compared to the ARIADNE Monte Carlo which
reasonably reproduces the data in each $(x,Q^2)$ interval.

The peak position of the distributions, \logxpmax, was evaluated by fitting
a Gaussian over a range $\pm1$ \logxp~unit around
the statistical mean. This fit was motivated by the MLLA prediction for the
form of the spectrum (appendix~A) which can be approximated by
a Gaussian distribution around the peak position at sufficiently high energies.
For consistency, the same fit was
performed on TASSO data \cite{tasso} at different centre of mass energies
and data from the OPAL experiment \cite{opal}. The \logxp~distributions
published in~\cite{tasso} and~\cite{opal} include charged
particles from $K^0_S$ and $\Lambda$ decays: Monte Carlo studies using JETSET
indicated that the effect of their inclusion on the peak position was less
than 0.5\%.
Results of the fits are shown in tables~\ref{table:ZEUSmax}
and~\ref{table:epem}.
Our fit to the $e^+e^-$ data
agrees, within statistical errors, with a similar
fit performed by the OPAL collaboration~\cite{kreutzmann}
\footnote{The OPAL analysis, using a modified Gaussian fit over a limited
range around the peak, was not repeated due to the limited
number of ZEUS data points. A modified Gaussian fit
over an extended range leads to consistent results for \logxpmax~for
the ZEUS data \cite{val}.}.

The ZEUS systematic error includes the contributions from refitting
the data with the systematic variations listed in section \ref{s:systematics}.
These were combined in quadrature with an estimate of the uncertainty on
the fit
obtained by varying the fit range and mean value by up to two bins \cite{val}.
The relative contribution to the systematic error from the fit variation
is $\simeq 40\%.$

\begin{table}[htb]
\begin{center}
\begin{tabular}{c r@{-}l r@{.}l c r@{.}l l} \hline
$x$ range                & \multicolumn{2}{c}{$Q^2$ range} &
\multicolumn{2}{c}{$Q$ mean}& \logxpmax $\pm$ stat $\pm$ sys &
\multicolumn{2}{c}{$\chi^2$/dof} \\
                         & \multicolumn{2}{c}{(GeV$^2$)} &
\multicolumn{2}{c}{(GeV)}   &    \\ \hline \hline
$0.6- 1.2\times 10^{-3}$  &  10&20                  & 3&8
& $1.49\pm 0.02\pm 0.06$   &  8&2/6\\ \\
$1.2- 2.4\times 10^{-3}$  &  10&20                  & 3&8
& $1.50\pm 0.03\pm 0.06$   &  4&7/6\\
                  &  20&40                  & 5&3
& $1.70\pm 0.03\pm 0.06$   & 14&4/8\\
                          &  40&80                  & 7&3
& $1.94\pm 0.06\pm 0.11$   &  4&4/6\\ \\
$2.4-10.0\times 10^{-3}$  &  20&40                  & 5&3
& $1.70\pm 0.03\pm 0.09$   & 12&5/6\\
                  &  40&80                  & 7&4
& $1.92\pm 0.03\pm 0.07$   &  5&8/6\\
	                  &  80&160                 &10&4                       &
$2.18\pm 0.06\pm 0.10$   &  3&2/6\\
 	                  & 160&320                 &14&5                       &
$2.25\pm 0.10\pm 0.23$   &  4&0/6\\ \\
$1.0- 5.0\times 10^{-2}$  & 320&640                 &20&4
& $2.78\pm 0.22\pm 0.26$   &  3&8/6\\
                          & 640&1280                &29&2
& $2.85\pm 0.21\pm 0.24$   &  5&6/8\\ \hline
\end{tabular}
\end{center}
\caption{\label{table:ZEUSmax}{\it Fitted values of \logxpmax~from
the ZEUS data.}}
\end{table}

\begin{table}[htb]
\begin{center}
\begin{tabular}{l c c r@{/}l} \hline
Experiment   & $\sqrt{s}$ (GeV) & \logxpmax $\pm$ stat   &
\multicolumn{2}{c}{$\chi^2$/dof} \\ \hline\hline
TASSO        & 14             & $2.356\pm 0.022$         & 3.9 &  7\\
             & 22             & $2.667\pm 0.024$         &  4.1 &  7\\
             & 35             & $3.020\pm 0.018$         & 15.3 &  7\\
             & 44             & $3.104\pm 0.021$         &  4.6 &  7\\
OPAL         & 91             & $3.594\pm 0.013$         & 19.1 & 17\\ \hline
\end{tabular}
\end{center}
\caption{\label{table:epem}{\it Fitted values of \logxpmax~from
the TASSO and OPAL data.}}
\end{table}

Figure~\ref{figure:slope} shows
the distribution of \logxpmax~as a function of $Q$
for the ZEUS data and of $\sqrt{s}$ for the $e^+e^-$ data.
Over the range shown
the peak moves from $\simeq$~1.5 to 2.8, equivalent to
the position of the maximum of the
corresponding momentum spectrum increasing from
$\simeq$~400 to 900~MeV.
The ZEUS data complement the $e^+e^-$ results.
%, extending the range in $Q$ to lower values. At higher $Q$
The ZEUS data points are
consistent with those from TASSO and a clear agreement in the rate of
growth of the ZEUS points with
the $e^+e^-$ data at higher $Q$ is observed.

The increase of \logxpmax~can be approximated phenomenologically
by the straight line fit
$$
\logxpmax\;=\; b\: \ln(Q)+c
\label{eqn:str_fit}
$$
also shown in figure~\ref{figure:slope}.
The values obtained from the fit to the ZEUS data are
$b=0.650 \pm 0.035{\rm (stat)}\pm 0.069{\rm (sys)}$ and
$c=0.626 \pm 0.059 \pm 0.129.$
The systematic errors are calculated by re-fitting the
\logxp~distributions obtained according
to the variations listed in section \ref{s:systematics} and combining the
deviations from the central value of the fit parameter, $b$ or $c$,
in quadrature.
The dominant error is from the correction of the data using the
HERWIG Monte Carlo model.
Removing the lowest two $Q$ points from the fit, where the range over which
the Gaussian is fitted extends well beyond the momentum range in which the MLLA
is valid (see section~\ref{sec:mlla}), leads to no statistically significant
deviation from the quoted values of $b$ and $c$.
The gradient
extracted from the OPAL and TASSO data is $b=0.653\pm0.012$ (with
$c=0.653\pm0.047$) which is consistent with the ZEUS result.
This value is consistent with that published
by OPAL, $b=0.637\pm 0.016$,
where the peak position was extracted using an
alternative method~\cite{opal}.
A consistent value of the gradient is therefore determined in DIS and
\ee~annihilation experiments.

Also shown is the statistical fit to the data
when $b=1$ ($c=0.054 \pm 0.012$) which would be the case if the QCD cascade
was of an incoherent nature, dominated by cylindrical phase space. (A
discussion of phase space  effects is given in~\cite{boudinov}.)
In such a case, the logarithmic particle momentum spectrum would be
peaked at a constant value of momentum, independent of $Q$.
The observed gradient is clearly inconsistent with $b=1$
and therefore inconsistent with cylindrical phase space.

In figure~\ref{figure:mcmodels}a the measurements are compared with the
results from the ARIADNE, HERWIG and MEPS Monte Carlo simulations.
These models describe the \logxp~distributions and the evolution of their
maximum as well as the
multiplicity distributions shown in section \ref{s:multi}.
As in the case of $<\!n_{ch}\!>$, the difference between
the coherent and incoherent growth of \logxpmax~is found to be
reduced when string fragmentation is used rather than
independent fragmentation~(figure~\ref{figure:mcmodels}b).
Again, the \logxp~distributions and the growth of \logxpmax~with $Q$ are
well-described by Monte Carlo models that include coherence in their
simulations, whilst those with an incoherent shower are less successful in
describing the data.

In contrast to the $<\!n_{ch}\!>$ measurements, the \logxpmax~values change
only slightly when only the rapidity gap events are used, with a maximum
negative shift of $\simeq 5\%$ in the lowest $(x, Q^2)$ interval.
Effects due to primary heavy-quark production were also found to be negligible.

The evolution of \logxpmax~seen in DIS agrees with the \ee~data.
The growth  is at a slower rate than that expected from
cylindrical phase space, indicative of coherence.
Monte Carlo models including coherence effects
describe the \logxpmax~values and the \logxp~distributions
better than those with an incoherent behaviour.
The rate of growth with increasing $Q$ of $<\!n_{ch}\!>$ and
\logxpmax~in the current region of the Breit frame agrees with \ee~data
in the region of overlap (i.e. above 10~GeV)
indicating that the fragmentation of quarks in DIS is similar
to that of quarks produced in \ee~annihilation.

\section{Comparisons with MLLA}
\label{sec:mlla}

In order to test the predictions of the MLLA (Modified Leading Log
Approximation) and
LPHD (Local Parton Hadron Duality) at different energies, fits to the
individual \logxp~distributions were performed in the chosen $(x,Q^2)$
intervals.
The evolution of these distributions with $Q$
is also predicted within the framework of the MLLA.

In the high energy limit, MLLA calculations
predict the momentum distribution of soft gluons, $\bar D^{\rm lim}$,
radiated by a quark
of energy $E~=~Q/2$ according to equation~(\ref{eq:mlla})
given in appendix~A.
$\bar D^{\rm lim}$ is known as the limiting spectrum.
Invoking LPHD, the \logxp~distribution of the hadrons is given by
\begin{eqnarray}
\frac{1}{\sigma}\frac{d\sigma}{d(\ln(1/x_{P}))} =
\kappa^{\rm ch}\bar D^{\rm lim}(\ln(1/x_{P}),Y) \label{eq:fiteqn}
\end{eqnarray}
where $Y = \ln (Q/2\Lambda)$.
The $\Lambda$ in the expression is an effective
scale parameter and not the standard QCD scale,
e.g. $\Lambda_{\rm \overline{MS}}$. This MLLA prediction is valid in the
range $0  \sleq \logxp \sleq Y$, a region where
pQCD can be applied.
The normalisation factor  $\kappa^{\rm ch}$ is
energy independent within the
framework of LPHD.
In addition, the charged multiplicity can be expressed as
\begin{eqnarray}
<\!n_{ch}\!> = \kappa^{\rm ch} {\cal N}(Y) \label{eq:nch}
\end{eqnarray}
where ${\cal N}(Y)$ is the integral of $\bar D^{\rm lim}$ given by
equation~(\ref{eq:ng}) in appendix~A and represents the soft gluon
multiplicity.

The evolution of \logxpmax~with $Q$ is a function of $\Lambda$ and
is given by
\begin{eqnarray}
\logxpmax= \frac{1}{2}Y + c_2\sqrt{Y} - c_2^2 + \left( {\cal
O}(Y^{-1/2})\right) \label{eq:evoleqn}
\end{eqnarray}
where $c_2$ is a constant calculated from the number of flavours and
colours. Taking the active number of flavours to be 3, the value of the
constant $c_2$ is 0.52. The term $ {\cal O}(Y^{-1/2})$ describes higher-order
effects.

The functional form of equation~(\ref{eq:fiteqn}) was fitted to the
\logxp~distributions.
This equation has two free parameters, $\kappa^{\rm ch}\ {\rm and}
\ \Lambda.$
The fit was performed for each $(x,Q^2)$ interval in the
range $0 < \logxp < \ln(Q/2p_0)$ where $p_0 = 400~{\rm MeV}$.
This restriction limits the comparisons of MLLA at the lower $Q$ values.
Due to the restricted
range of the fit very little of the peak region of the lowest $Q$ value
is included: these intervals are therefore excluded from the
following fits. The results of these fits are shown in
table~\ref{table:mllaxp}.
\begin{table}[bht]
\begin{center}
\begin{tabular}{c r@{-}l r@{.}l c c } \hline
$x$ range                & \multicolumn{2}{c}{$Q^2$ range}
&\multicolumn{2}{c}{$Q$ mean}&$\Lambda \pm$ stat & $\kappa^{\rm ch} \pm$ stat
\\
                         & \multicolumn{2}{c}{(GeV$^2$)}&
\multicolumn{2}{c}{(GeV)} &(MeV)   &    \\ \hline \hline
$1.2- 2.4\times 10^{-3}$  &  20&40  &5&3                & $330\pm 29$ &
$1.25\pm 0.05$ \\
                          &  40&80  &7&3                & $323\pm 20$ &
$1.28\pm 0.07$ \\ \\
$2.4-10.0\times 10^{-3}$  &  20&40  &5&3                & $313\pm 28$ &
$1.21\pm 0.04 $\\
                  &  40&80 &7&4                         & $296\pm 19$ &
$1.30\pm 0.05$ \\
	                  &  80&160 &10&4                & $296\pm 28$ & $1.25\pm 0.06
$\\
 	                  & 160&320 &14&5                & $311\pm 48$ & $1.31\pm
0.10 $\\ \\
$1.0- 5.0\times 10^{-2}$  & 320&640 &20&4                & $210\pm 54$ &
$1.14\pm 0.10 $\\
                          & 640&1280 &29&2               & $266\pm 81$ &
$1.18\pm 0.16 $\\ \hline
\end{tabular}
\end{center}
\caption{\label{table:mllaxp}{\it Fitted values of $\Lambda$ and $\kappa^{\rm
ch}$.}}
\end{table}

Approximately the same value of $\Lambda$ is found at all values of $Q.$
The values of $\Lambda$ are combined to give a weighted mean value of
$$\Lambda = 306 \pm 10 {\rm \ (stat)} \pm 30 {\rm \ (sys)} {\ \rm MeV}.$$
The fits in each interval are also consistent with a single value of
$\kappa^{\rm ch}$
with a mean value of $\kappa^{\rm ch}$ given by
$$\kappa^{\rm ch} = 1.25 \pm 0.02 {\rm \ (stat)} \pm 0.09 {\rm \ (sys)}.$$
The experimental systematic errors are calculated by re-fitting the
\logxp~distributions obtained according
to the variations listed in section \ref{s:systematics} and combining the
deviations from the central values of the fit parameters,
$\Lambda$ and $\kappa^{\rm ch}$, in quadrature.
The dominant error is from the correction of the data using the
HERWIG Monte Carlo model.
The absolute value of the
correlation coefficient between $\Lambda$ and $\kappa^{\rm ch}$ from the
fits was typically $\sim 0.35.$
Equation~(\ref{eq:fiteqn}) was re-fitted
with $p_0$ varying from 200 to 500 MeV, yielding consistent results within
the quoted systematic errors.
Similarly, this equation was re-fitted
with $\Lambda$ fixed at its mean value and $\kappa^{\rm ch}$ as a
free parameter. The values of $\kappa^{\rm ch}$ remain the same within
the statistical accuracy of the fit with little degradation in the
$\chi^2$.

The results of the fits to the MLLA limiting spectrum
using the mean values of $\Lambda$ and $\kappa^{\rm ch}$ above are shown in
figure~\ref{figure:mllaevol}. The data
are reasonably well described by the limiting spectrum with single values
of $\Lambda$ and $\kappa^{\rm ch}$ over a large range in $Q$.
The overall $\chi^2/\!{\rm dof}$ for all the ZEUS data is $152/106.$
At higher values
of \logxp, beyond the range of the fit, the MLLA spectrum underestimates the
multiplicity.

The values measured by OPAL, fitting their data over a limited momentum
range, were \linebreak
$\Lambda =253\pm 30$ MeV and $\kappa^{\rm ch} = 1.44\pm 0.01.$
In order to compare with the ZEUS measurement, their value of $\kappa^{ch}$
has been divided by a colour factor of $4/9$
and by a factor of 2 to account for the quark and anti-quark pair,
factors which OPAL absorbed into their value quoted in~\cite{opal}.
The inclusion of decay-particles from $K^0_S {\rm\ and\ } \Lambda$
in the OPAL data contributes $10\%$ to the multiplicity~\cite{nch:opl}
and can account for the higher value of $\kappa^{\rm ch}$.
The value of $\Lambda$ is also sensitive to this but to a lesser degree.
Monte Carlo studies using JETSET indicated that the effect of their
inclusion on the value of $\Lambda$ was less than 10~MeV.

In order to check the consistency of $\kappa^{\rm ch}$,
equation~(\ref{eq:nch}) was fitted to $<\!n_{ch}\!>$ as a function of $Q$.
Taking the value of $\Lambda$ extracted from the
individual \logxp~distributions, $\Lambda = 306$ MeV, the fitted value of
$\kappa^{\rm ch} = 1.78\pm 0.02\,{(\rm stat)}$ was determined.
This value of $\kappa^{\rm ch}$ is
significantly higher than that extracted from the \logxp~distributions.
Equation~(\ref{eq:nch}) is derived from the integration of
equation~(\ref{eq:fiteqn}), this expression underestimates the data
at large \logxp~(see figure~\ref{figure:mllaevol}).
At the relatively low $Q$ values of the ZEUS data, $ Q \sleq 7$~GeV,
this accounts for a significant fraction of the total multiplicity
where the MLLA prediction, taken outside its range of validity, underestimates
the data.
The underestimate of the predicted parton multiplicity by the MLLA is
therefore compensated by an increase in the value of $\kappa^{\rm ch}.$

In order to check the consistency of $\Lambda$,
the evolution of \logxpmax~was investigated. The value of
$\Lambda$ hardly influences the slope of \logxpmax~in
equation~(\ref{eq:evoleqn}), but gives the absolute value of
\logxpmax~at a given $Q$ value. This is highly correlated with
the higher-order term ${\cal O}(Y^{-1/2})$ in equation~(\ref{eq:evoleqn})
and therefore this term is neglected in the following fits.
The Gaussian fits which determine the \logxpmax~values described in
section~\ref{s:logxp} extend to momentum values
lower than the measured $\Lambda.$
In order to compare with MLLA calculations,
the peak position of the \logxp~distribution was extracted again,
using a Gaussian symmetric around the estimated peak position.
The range of the fit was from $(2L_0 - Y)$ to $Y$
where $L_0$ is the estimated peak position, calculated from
equation~(\ref{eq:evoleqn}) neglecting the
${\cal O}(Y^{-1/2})$ term and assuming $Y=\ln(Q/2p_0),$
where $p_0 = 306$ MeV.
A value
of $\Lambda = 284\pm 8{(\rm stat)}$ MeV is found,
consistent with the value extracted from the fit of the individual plots.
This compares to a value of $\Lambda = 212\pm 20$ MeV from the evolution
of the TASSO and OPAL data as extracted by OPAL.
(OPAL fit equation~(\ref{eq:evoleqn}) with
$c_2$ as a free parameter; a high correlation was found between $c_2$ and
$\Lambda$ when fitting the ZEUS data.)

Although the growth of the multiplicity cannot be consistently
described by the MLLA and LPHD,
a single value of $\Lambda$ and $\kappa^{\rm ch}$
can describe the \logxp~distributions over a range of
$Q$ values from $\sim 5\ {\rm to\ } 30$~GeV.
The values obtained are consistent with those extracted at
$\sqrt{s} = 91$~GeV by the OPAL experiment.
The evolution of \logxpmax~with $Q$ also gives a consistent value
of $\Lambda$, again compatible with the OPAL and TASSO results.

\section{Conclusions}

Charged particle distributions have been measured
in the current region of the Breit frame in DIS over a wide range
of $Q$ values.
In the framework of the Monte Carlo models, the best description of the
data is achieved by those incorporating coherence effects where
the production of soft particles and the growth of the mean
multiplicity is suppressed relative to the incoherent case.
Modified Leading Log Approximation (MLLA)
calculations have been shown to describe the
\logxp~distributions in the range
$5~\sleq~Q~\sleq~30$~GeV using single values of $\kappa^{\rm ch}$, which
relates the observed hadron distributions to the calculated parton
distributions, and of $\Lambda$, an effective scale parameter of the QCD
calculations.

A comparison of these charged particle distributions
with those from \ee~annihilation
experiments has been performed.
For $Q > 10$~GeV a growth in $<\!n_{ch}\!>$ is observed with $Q$ which is
similar to the growth in \ee~experiments as a function of $\sqrt{s}$. The
evolution of \logxpmax~has been measured and is consistent
with that observed in \ee~experiments.
The fragmentation
properties of the struck quark from the proton in DIS
that have been studied here
are similar to those from quarks created in \ee~annihilation.
The observed charged particle spectra are therefore consistent
with the universality of quark fragmentation.

\section*{Acknowledgements}

The experiment was made possible by the inventiveness and the diligent
efforts of the HERA machine group who continued to run HERA most
efficiently during 1993.

The design, construction and installation of the ZEUS detector has
been made possible by the ingenuity and dedicated effort of many people
from inside DESY and from the home institutes who are not listed as authors.
Their contributions are acknowledged with great appreciation.

The strong support and encouragement of the DESY Directorate
has been invaluable.

We also gratefully acknowledge the support of the DESY computing and network
services.

We would like to thank V. Khoze and B. Webber for valuable discussions.

\newpage
\section*{Appendix A}
\subsection*{MLLA formul\ae}

QCD predicts the form of the soft gluon momentum distribution at a
fixed energy and describes the energy evolution of the spectrum, including
the multiplicity and the peak maximum.

In the high energy limit, MLLA (pQCD) calculations
predict~\cite{MLLA} the momentum distribution of soft gluons from a quark
with energy $E$ according to the following formula, in which $x_P$
is the momentum scaled with $E$,
\begin{eqnarray}
\bar D^{\rm lim}(\ln(1/x_{P}),Y)& =& \frac{4C_f}{b}\Gamma (B)
\int_{-\pi/2}^{\pi/2} \frac{d\tau}{\pi}e^{-B\alpha}
\left[ \frac{\cosh \alpha +(1-2\zeta)\sinh \alpha
}{\frac{4N_c}{b}Y\frac{\alpha}{\sinh\alpha}} \right]^{B/2} \nonumber  \\
& & \cdot I_B\left( \sqrt{\frac{16N_c}{b}Y\frac{\alpha}{\sinh\alpha}
\left[\cosh \alpha+(1-2\zeta)\sinh \alpha \right]}~\right)
\label{eq:mlla}
\end{eqnarray}
where $C_f,\ b$ and~$B$ are constants based on the number of
flavours and colours.
For the number of flavours, $N_f = 3$, and the number of colours,
$N_c=3$, the constants $C_f,\ b\
{\rm and \ } B$ are equal to $4/3,\ 9,\ {\rm and\ }1.247$
respectively.
$Y = \ln (E/\Lambda)$,
where $\Lambda$ is an effective
scale parameter.
For the case of DIS, in the current
fragmentation region, $E= Q/2$ and for \ee,~$E=\sqrt{s}/2.$
$\Gamma$ denotes the Gamma
function and $I_B$ is a modified Bessel function of order $B$.
The variable $\alpha$ is given by
$\alpha = \alpha_0 + i\tau$, where $\alpha_0 $ is determined by
\begin{eqnarray}
\tanh \alpha_0 = 2\zeta - 1 \nonumber
\end{eqnarray}
and $\zeta = 1 - \ln(1/x_{P})/Y.$
The integral, which is a representation of the confluent
hypergeometric function, is performed over $\tau$.
$\bar D^{\rm lim}$ is known as the limiting
spectrum which is expressed in a form
that is suitable for numerical integration.

{}From the limiting spectrum
the soft gluon multiplicity can be calculated as
\begin{eqnarray}
{\cal N}(Y) = \frac{C_f}{N_c}\Gamma (B)\left(\frac{z}{2}\right)^{-B+1}
I_{B+1}(z)
\label{eq:ng}
\end{eqnarray}
where
$ z = \sqrt{\frac{16N_c}{b}Y}.$

%
%--------- REFERENCES -------------
%

%
\newpage
%----- Figures -----------
\begin{figure}[tp]
\centerline{\psfig{figure=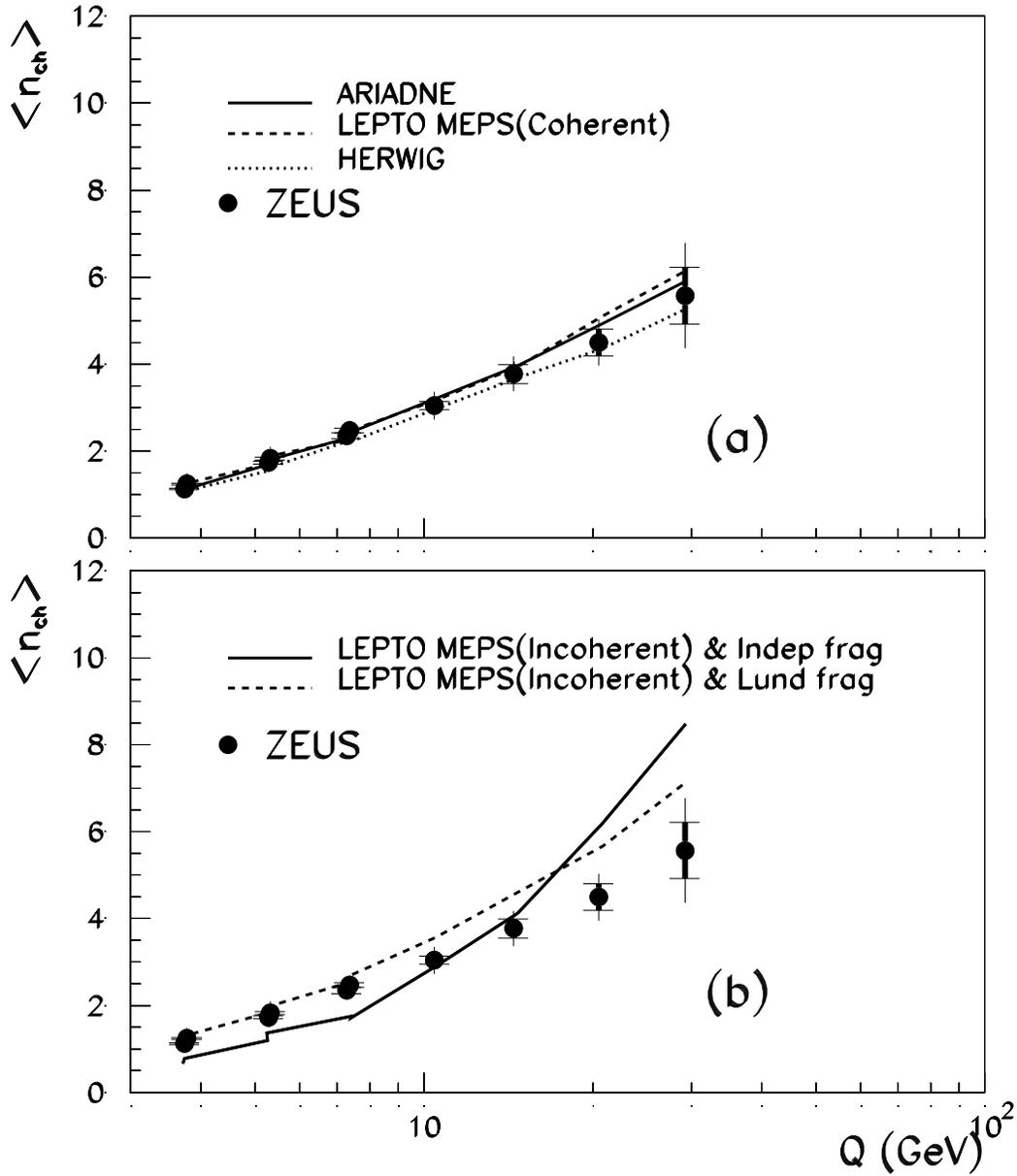,width=18cm}}
\caption{{\it $<\!n_{ch}\!>$ as a function of $Q$
for the $x$-intervals shown in table~1. The ZEUS data
are compared to various Monte Carlo models.
The data in figures (a) and (b) are the same.
The discontinuities in the Monte Carlo curves correspond to
predictions for two different $x$ values at fixed $Q$.
The line connects the lower-$x$ to the higher-$x$ point with increasing $Q$.
}}
\label{figure:meannch}
\end{figure}
\begin{figure}[tp]
\centerline{\psfig{figure=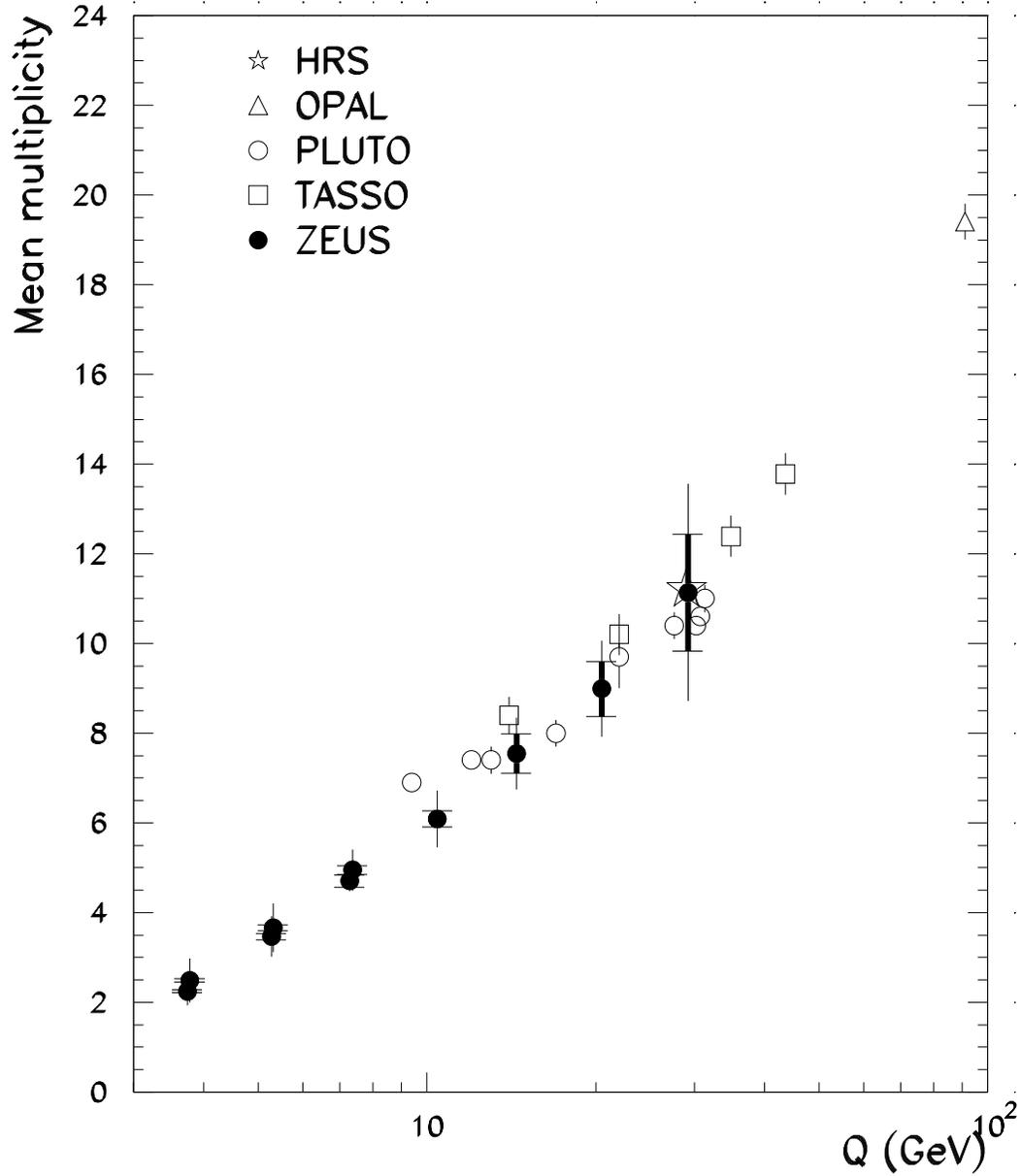,width=18cm}}
\caption{{\it Mean charged multiplicity as a function of $Q$.
Twice the measured ZEUS multiplicity, $2\cdot <\!n_{ch}\!>$,  is compared
to $<\!n_{ch}\!>$ results from HRS, OPAL, PLUTO and TASSO.
All data were corrected for $K^{0}_S$ and $\Lambda$ decays.
}}
\label{figure:nchepem}
\end{figure}
\begin{figure}[tp]
\centerline{\psfig{figure=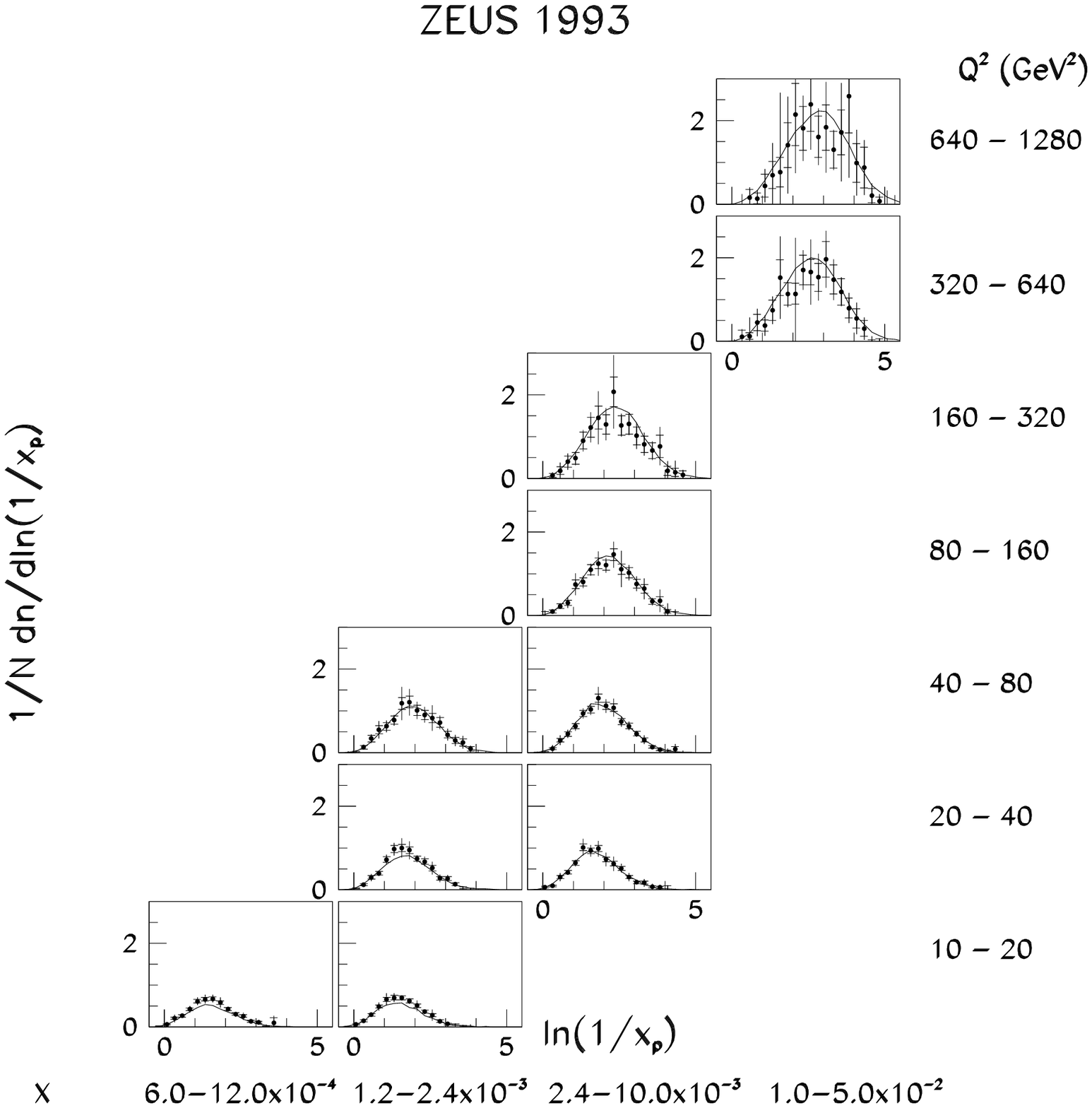,width=18cm}}
\caption{{\it
\logxp~distributions in the current region of the Breit frame
as a function of $(x, Q^2)$. Statistical errors are indicated
by the inner error bar bounded by the horizontal bars. The outer
error bars show the statistical and systematic errors added in quadrature.
The ZEUS data are compared to the predictions of the ARIADNE Monte
Carlo.}}
\label{figure:corr_data}
\end{figure}
\begin{figure}[tp]
\centerline{\psfig{figure=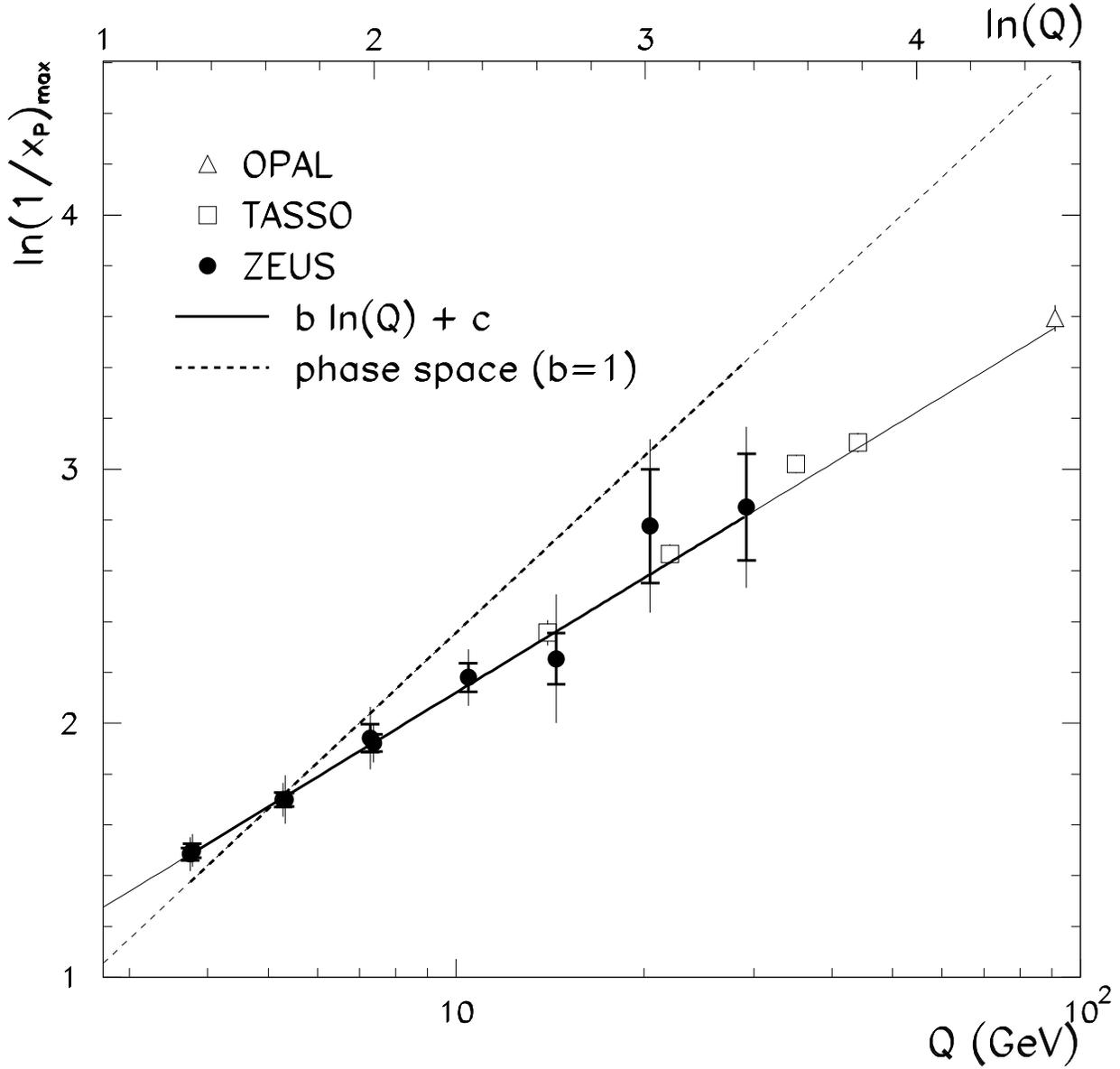,width=18cm}}
\caption{{\it \logxpmax~as a function of $Q$
for the $x$-intervals shown in table~1. The ZEUS data are compared
to results from OPAL and TASSO.
A straight line fit of the form $\logxpmax~=~b~ln(Q)~+~c$
to the ZEUS \logxpmax~values is indicated as well as the line
corresponding to $b~=~1$, discussed in the text.}}
\label{figure:slope}
\end{figure}
\begin{figure}[tp]
\centerline{\psfig{figure=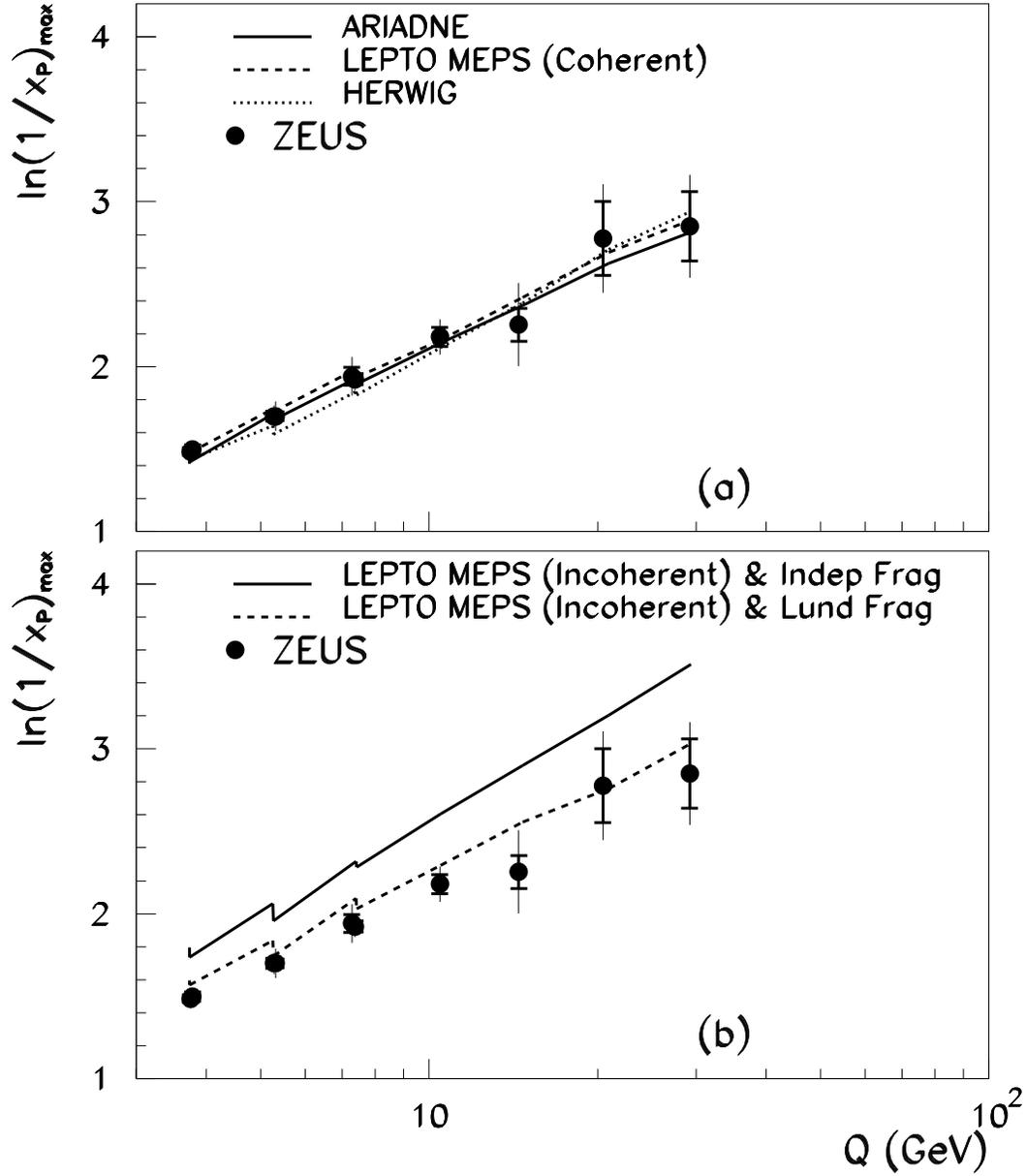,width=18cm}}
\caption{{\it \logxpmax~as a function of $Q$. The ZEUS data
are compared to various Monte Carlo models.
The data in figures (a) and (b) are the same.
The discontinuities in the Monte Carlo curves correspond to
predictions for two different $x$ values at fixed $Q$.
}}
\label{figure:mcmodels}
\end{figure}
\begin{figure}[tp]
\centerline{\psfig{figure=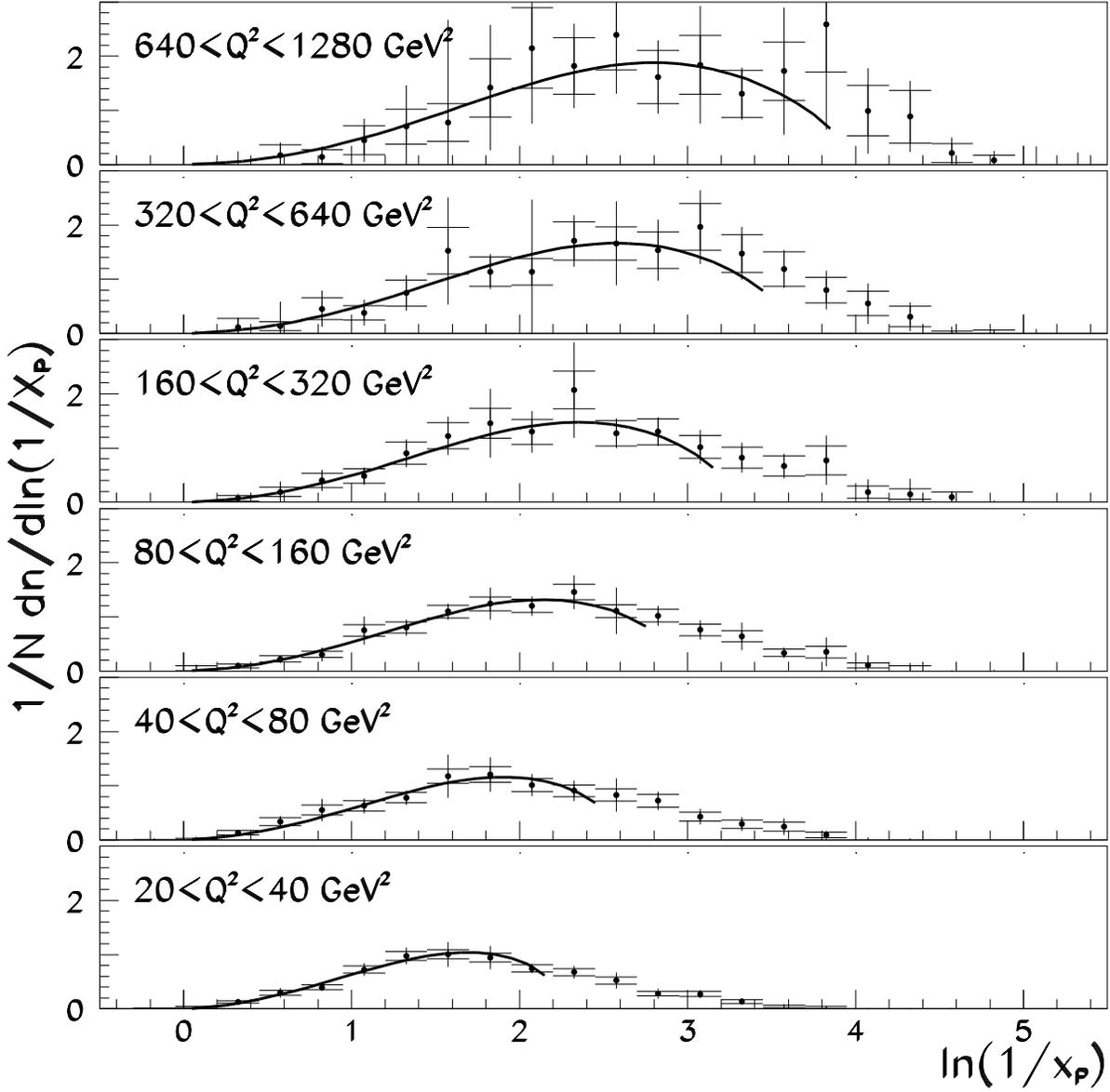,width=18cm}}
\caption{{\it \logxp~distributions
in the current region of the Breit frame
as a function of $Q^2$.
The ZEUS data are compared to the MLLA limiting
spectrum in the range $0 < \logxp < \ln(Q/2\Lambda)$
with $\Lambda$ = 306MeV and $\kappa^{\rm ch}$= 1.25.
For clarity, only the
lower $x$ range is shown in the two lowest $Q^2$ intervals.}}
\label{figure:mllaevol}
\end{figure}
\end{document}